\documentclass[useAMS,usenatbib]{mn2e}
\usepackage{graphicx}
\usepackage{latexsym}

\title{Ionisation-induced star formation II: External irradiation of a turbulent molecular cloud}

\author[J. E. Dale, P. C. Clark, I.A. Bonnell]{J. E. Dale$^{1}$\thanks{E-mail: Jim.Dale@astro.le.ac.uk (JED)},
P. C. Clark$^{2}$,I. A. Bonnell$^{3}$\\
$^{1}$Department of Physics and Astronomy, University of Leicester, University Road, Leicester, LE1 7RH\\
$^{2}$Institut f\"ur Theoretishce Astrophysik, Universit\"at Heidelberg, Albert--\"Uberle--Str. 2, Heidelberg, Germany \\
$^{3}$Department of Physics and Astronomy, University of St Andrews, North Haugh, St Andrews, Fife KY16 9SS}

\begin{document}

\pagerange{\pageref{firstpage}--\pageref{lastpage}} \pubyear{2006}

\maketitle

\label{firstpage}

\def\mnras{MNRAS}
\def\apj{ApJ}
\def\aj{AJ}
\def\aap{A\&A}
\def\apjl{ApJL}
\def\apjs{ApJS}
 
\begin{abstract}
In this paper, we examine numerically the difference between triggered and revealed star formation. We present Smoothed Particle Hydrodynamics (SPH) simulations of the impact on a turbulent $10^{4}$ M$_{\odot}$ molecular cloud of irradiation by an external source of ionising photons. In particular, using a control model, we investigate the triggering of star formation within the cloud. We find that, although feedback has a dramatic effect on the morphology of our model cloud, its impact on star formation is relatively minor. We show that external irradiation has both positive and negative effects, accelerating the formation of some objects, delaying the formation of others, and inducing the formation of some that would not otherwise have formed. Overall, the calculation in which feedback is included forms nearly twice as many objects over a period of $\sim0.5$ freefall times ($\sim2.4$ Myr), resulting in a star--formation efficiency approximately one third higher ($\sim4\%$ as opposed to $\sim3\%$ at this epoch) as in the control run in which feedback is absent. Unfortunately, there appear to be no observable characteristics which could be used to differentiate objects whose formation was triggered from those which were forming anyway and which were simply revealed by the effects of radiation, although this could be an effect of poor statistics.
\end{abstract}

 
\section{Introduction}
Stars form in molecular clouds, but the question of what \textit{causes} a molecular cloud to form stars is an intriguing one. Numerous means of triggering star formation have been proposed. On the largest scales, collisions or tidal interactions between galaxies (e.g. \cite{2001AJ....121..768D}), passage of galactic spiral arms through the ISM (e.g. \cite{1990ApJ...349..497C}, \cite{2002MNRAS.337.1113S}, \cite{2006MNRAS.365...37B}) and collisions between molecular clouds (e.g. \cite{1994A&A...290..421W}) have all been put forward. Alternatively, observations suggest Giant Molecular Clouds may be able to form stars spontaneously \citep{2000ApJ...530..277E}. In this paper we concentrate on processes acting on smaller lengthscales than any of those listed above; processes driven by stars themselves. We investigate whether stellar feedback can trigger star formation and also examine the extent to which feedback simply reveals stars forming spontaneously.\\
\indent The Jeans mass, $M_{J}$ determines the mass scale on which gas fragments under the influence of gravity (e.g. \cite{1985MNRAS.214..379L}). The Jeans mass depends on temperature and density such that $M_{J}\propto T^{1.5}\rho^{-0.5}$. All the putative triggering mechanisms listed above rely on encouraging fragmentation and star formation by increasing the density of molecular gas while leaving its temperature roughly constant. Most achieve this by driving isothermal shocks into the gas. The localised injection of mechanical or thermal energy into the ISM by massive stars via photoionization, winds or supernova explosions is an obvious way of doing this. The triggering of star formation by stars themselves would make star formation a self--propagating process (e.g. \cite{1977ApJ...214..725E}, \cite{1981ApJ...249...93S}, \cite{1983ApJ...265..202S}), spreading through galaxies like a viral infection.\\
\indent Triggering of star formation by feedback from O--type stars has been studied observationally and theoretically by several authors. \cite{1994A&A...290..421W} examined the generic problem of the fragmentation of a shocked shell driven into a uniform medium by an expanding HII region or stellar wind bubble and derived expressions for the time and radius at which the shell fragments and for the mass of the fragments produced, concluding that this process is likely to give to birth to relatively massive stars. \cite{2005A&A...433..565D} have found evidence of triggered star formation around HII regions closely matching the scenario presented by \cite{1994A&A...290..421W} (we perform numerical simulations of this scenario in a companion paper). In addition \cite{1977ApJ...217..473H} have observed apparent rings of star formation around supernova remnants. \cite{1977ApJ...214..725E} considered the more difficult problem of a molecular cloud irradiated by an external OB association. In this picture, stellar radiation ionises the skin of the cloud which `boils off', exerting a reaction force on the cloud as it does so, and driving a shock into the remaining neutral gas. \cite{2005A&A...433..955C} recently made an observational study of the RCW 108 molecular cloud in which they claim to observe this process.\\
\indent These observational and theoretical studies of self--propagating star formation can take us a long way towards understanding the process, but suffer from several drawbacks. Analytical studies can only consider uniform molecular clouds and can only indicate when and where star formation is likely to occur and the average properties of the stars formed. Observational studies are hampered by the exceedingly difficult task of deciding whether or not a given star was induced to form, whether it was going to form anyway, or whether it had \textit{already} formed and has merely been revealed by the passage of a shock (\cite{2005ASPC..341..107H}, \cite{2003ApJ...595..900K}). Distinction between the first two of these possibilities is rarely made, since it is all but impossible to distinguish between them observationally. Such studies usually rely on geometrical association of YSOs with the boundaries of HII regions or SNRs. Additional checks are provided by comparing the stellar ages to the age of the HII region or SNR and, in the former case, the spatial association of the YSO with bright ionised gas or cometary globules. \cite{2005ASPC..341..107H} in particular suggest that the degree to which star formation in a given region is triggered may be assessed by looking for geometrical correlations between low mass stars and high mass stars, and between low mass stars and gas compressed by HII regions (compare their Figures 8 and 9).\\
\indent Hydrodynamical simulations offer ways around some of these drawbacks. Realistic non--uniform molecular clouds with turbulent velocity fields can be modelled. Whereas an observer has only a single snapshot of a molecular cloud to study, a computational astrophysicist has the benefit of hindsight. In addition, numerical workers can perform experiments, in which the reaction of a model cloud to the influence of feedback can be compared to the behaviour of a `control' cloud. We report the results of such an experiment here.\\
\indent This work is intended to answer three simple questions:\\
(1) Can ionising feedback from O--stars on a GMC accelerate the formation of stars that the cloud would have formed in the absence of feedback?\\
(2) Conversely, can ionising feedback delay or prevent the formation of stars that a GMC would otherwise have formed?\\
(3) Can ionising feedback cause a GMC to form stars that it would not otherwise have formed?\\
We refer to the accelerated formation of stars that would form anyway as `weak triggering' and to the formation of stars that would not otherwise form as `strong triggering'. We feel this distinction is important and, at present, it is only through numerical simulations that it can be made apparent.\\
\indent In Section 2 we present our numerical methods. We discuss our choice of initial conditions in Section 3 and present our results in Section 4. In Section 5 we discuss the implications of our results for the study of triggered star formation, particularly from an observational perspective. Our conclusions are presented in Section 6.\\

\section{Numerical Methods}
The SPH code used in these calculations is described in \cite{1995MNRAS.277..362B}, although we have modified the code to simulate the presence of point sources of ionising radiation \citep{2005MNRAS.358..291D}. Our method is essentially a `Str\"omgren volume' method. Every SPH particle is examined to determine the flux of ionising photons reaching that particle. The particles are first sorted by increasing radius from the source. A vector is drawn between each target particle and the radiation source. Other SPH particles near this line--of--sight vector are located using an algorithm similar to that presented in \cite{2000MNRAS.315..713K} and their densities are used to form a discretised radial density profile along the line--of--sight vector. The ionising photon flux $S$ per unit solid angle reaching the particle is then determined from the $i$ positions $r_{i}$ and densities $\rho_{i}$ of the selected particles along the vector as
\begin{eqnarray}
S=\frac{L_{*}}{4\pi}-\sum_{i}\alpha_{B}\frac{(\rho_{i}+\rho_{i-1})}{2}r_{i-1}^{2}\Delta r,\\
\textrm{where }\Delta r=(r_{i}-r_{i-1}),
\end{eqnarray}
and $\alpha_{B}$ is a recombination coefficient, defined such that the number of recombinations to all atomic hydrogen states except the ground state per unit volume per unit time is given by $\alpha_{B}n^{2}$.\\
\indent If $S<0$, this implies that all ionising photons emitted in the direction of the target particle are consumed by recombinations before reaching it. If this target particle is neutral, it remains so. If it is ionised, the number of recombinations occurring within the particle during the timestep is calculated and its ionisation fraction decreased. Ionised particles which become fully neutral in this way are allowed to cool using a cooling curve obtained from \cite{1993A&A...273..318S}.\\
\indent If $S>0$, ionising photons are reaching the target particle. If it is already ionised, it is assumed to remain so. If it is neutral and the photon flux is sufficient to ionise it during the current timestep, it is ionised and its temperature raised to $10^{4}$ K. If the flux is too low to ionise the particle immediately, the photons reaching it are `stored', so that the particle may accumulate sufficient photons in the future to become ionised. This mechanism is important, since particles far from the radiation source may be formally inside the Str\"omgren volume, but receiving such a small photon flux that they take a dynamically--significant time to become ionised. We tested the photon--storage algorithm by modelling the approach of an ionisation front to the Str\"omgren radius in a uniform medium.\\
\indent It is usual to take \textit{secondary} ionising photons (i.e. those produced by recombinations directly to the ground state) to be absorbed immediately very close to their site of emission -- this is referred to as the `on--the--spot' (OTS) approximation. This approximation is easily made by modifying the recombination coefficient $\alpha$ so that recombinations directly to the ground state are ignored, yielding the `modified recombination coefficient' $\alpha_{B}$. We have made this approximation in this work, although we note that it may not be strictly valid in the geometry employed in these calculations. The OTS approximation relies on the ionised gas being sufficiently dense that secondary ionising photons are very likely to be absorbed within the HII region. In our calculations, we are irradiating the skin of a molecular cloud, so that ionised gas `evaporates' off the surface of the cloud. The ionised gas may then be of low density and it is possible that some fraction of the secondary photons may escape, violating the OTS approximation. We note, however, that $\alpha\approx4/3\alpha_{B}$. This implies that taking the other extreme and assuming that all the secondary photons are \textit{lost}, and that recombinations direct to the ground state constitute an additional photon sink,  only decreases the flux reaching the ionisation front by $33\%$. Given that the gas density just beyond the ionisation front is very high, this error in flux is likely to generate a very small error in the location of the ionisation front in any given direction. In any case, the change in recombination coefficient can be offset by a change in the source luminosity. Assuming that all the secondary photons are lost is equivalent to requiring a source $33\%$ brighter. For this reason, we make use of the OTS approximation in our calculations for simplicity.\\
\indent We also neglect the likely presence of a photo--dissociated region (a PDR) beyond the ionisation front, in which sub--Lyman photons would dissociate and heat the molecular gas. The problem of the simultaneous propagation of an ionisation front and a photodissociation front into an externally--irradiated cloud has been considered analytically by \cite{1996ApJ...458..222B}. They considered a problem similar to (although simpler than) that presented here and identified two possibilities: either the ionisation front subsumes the photodissociation front, in which case there is no appreciable PDR, or the photodissociation front can outrun the ionisation front and a PDR grows between the two fronts. Which of these possibilities occurs depends on the ratio $S_{Ly}/S_{FUV}$ of Lyman--continuum to far--UV (i.e. $91.2-111$ nm) photons and the Lyman--continuum optical depth $\tau_{Ly}$ between the source and the ionisation front (see their Equation 19). For a source with $S_{Ly}=10^{49}$s$^{-1}$ (as we willuse), $S_{Ly}/S_{FUV}\approx1$. In the very early stages of our simulation before the photoevaporation flow has established itself, $\tau_{Ly}$ can be large, but during most of the simulation it is low ($\sim0.05$), implying (again from Equation 19 of \cite{1996ApJ...458..222B}) that the photodissociation front in this case will not be able to outrun the ionisation front and that the PDR would consequently be narrow or non--existent.\\
\indent The star formation process in the code is modelled by sink particles which are allowed to form from clumps of gas particles if the density of the clump exceeds a threshold (in this case $10^{2}\times$ the mean density of the cloud) provided that the mass of the clump exceeds the local Jeans mass and that the clump is contracting. Once formed, a sink particle is given an accretion radius. Gas particles straying inside the accretion radius may be accreted by the sink particle if they pass a series of tests \citep{1995MNRAS.277..362B}. In these calculations, the accretion radius assigned to sink particles is large, $\sim1.7\times10^4$ au, so the sink particles should not be regarded as individual stars, but as multiple systems or star--forming cores. We assume that all the mass accreted by these cores is accreted by the one of the notional stars that the core contains, so that the total mass contained in all such cores can be taken as the total stellar mass. The sink particles therefore trace the spatial distribution and efficiency of star formation, but cannot be used to construct stellar mass functions (in contrast to \cite{2003MNRAS.343..413B}).\\
\indent The SPH code used in these simulations makes use of individual particle timesteps to increase efficiency. We found that, in this problem and in common with \cite{2005MNRAS.358..291D}, it was necessary to update the Str\"omgren volume on the shortest dynmical timestep being used by the code. The problem in question is highly dynamic and the gas as seen from the radiation source is very anisotropic. As a result, we found that varying the timestep on which the Str\"omgren volume is updated produces small variations in the quantity of ionised gas existing at a given time. These variations tend to be cumulative with time, in the sense that the difference betweem the ionisation fractions of two runs with different ionisation timesteps tend to increase with time, and can reach values of a few percent after half a freefall time. Although this variation is superficially small and produces only minor changes in global quanities such as overall star--formation efficiency, it can have an effect on the star--formation history of the simulation. This is not surprising, given that star formation in turbulent gas is a stochastic process, and that it is a relatively small quantity of ionised gas near the ionisation front that affects the star--formation. We found that these small changes in the quantity of ionised gas could alter the order in which objects form, affect the accretion history of objects and, in one extreme case, determine whether the formation of an object was prevented by ionisation or not (the total number of objects in existence at any given time could vary by as much as $10\%$). These alterations do not converge in any sense and instead appear to be stochastic. Although these variations have little effect on our central conclusions, this paper is concerned with examining the history of individual objects. We therefore chose to update the Str\"omgren volume every time the positions of  \textit{any} gas particles were updated (i.e. on the shortest dynamical timestep being used at any given time) so that the motion of the densest gas (that which is most likely to be forming stars) is followed self--consistently by the ionisation code. The overhead incurred by the ionisation algorithm is then quite severe and comes to dominate the runtime in later stages of the calculation. As a consequence, our feedback calculation took approximately nine weeks to run on a $4\times2.4$ GHz SUN v40z machine, as opposed to only one week for the control run.\\

\section{Initial conditions: turbulent molecular clouds}
\begin{figure}
\includegraphics[width=0.5\textwidth]{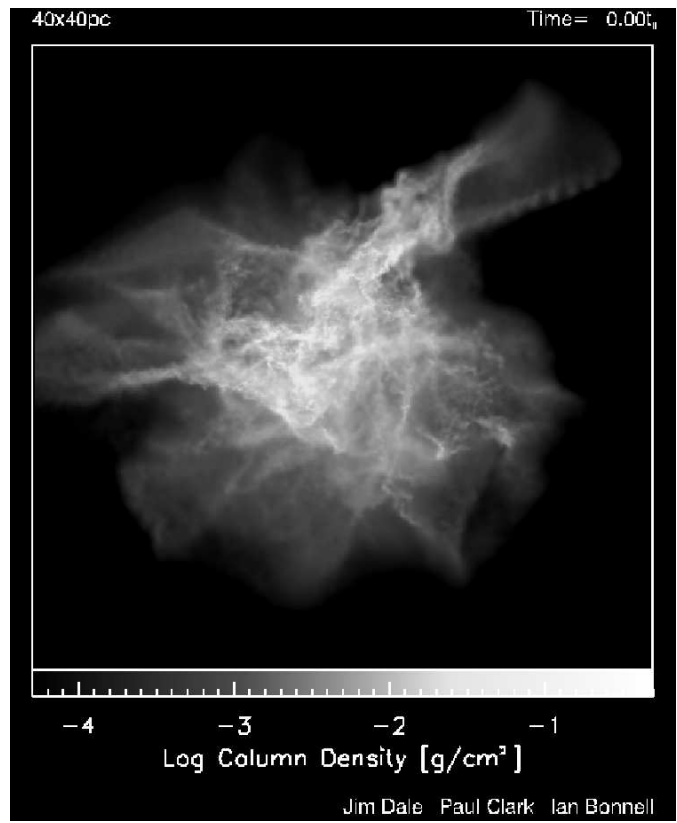}
\caption{Column density map, viewed along the z--axis, of the initial conditions of our simulations.}
\label{fig:init_snap}
\end{figure}

The initial conditions for our simulations are similar to those used by \cite{2005MNRAS.359..809C} to study star formation in unbound molecular clouds. Using $10^{6}$ particles, we constructed a uniform spherical cloud $10$ pc in radius with a mass of $10^{4}$ M$_{\odot}$ and a turbulent velocity field with an initial Mach number of $10$ and an energy $E_{turb}=2|E_{grav}|$, where $E_{grav}$ is the cloud's gravitational potential energy. The cloud is therefore globally unbound, but \cite{2005MNRAS.359..809C} showed that such clouds are still able to forms stars, albeit with low efficiency ($\sim10\%$ after $\sim2$ freefall times, at which point they assumed that the cloud was likely to suffer an internal supernova, destroying it and terminating star formation). Since star formation in this cloud is expected to be inefficient, it should reveal very clearly the effect on star formation of an external ionising radiation field. The turbulent velocity field generates complex structure within the cloud. The cloud's unbound state causes it to globally expand in size, although the central regions contract and become self--gravitating, allowing it to form dense star--forming cores. Clearly, the cloud is likely to form massive stars which would have strong ionizing radiation fields of their own, but we neglect this possibility here, as we are only interested in the effect of an external radiation field.\\
\indent We allowed the cloud to evolve undisturbed until just before it formed its first star--forming core after $\sim0.23$ freefall times ($\sim1.1$ Myr). Figure \ref{fig:init_snap} shows a column--density map projected along the $z$--axis of the cloud at this stage. The cloud now has a half--mass radius of $\sim10$pc although it continues to expand during our simulations due to its super--virial state. Its mean density is low and its mean initial Jeans mass is $\sim60$ M$_{\odot}$, so that the cloud contains $\sim170$ Jeans masses. We made two copies of the cloud in this state and reset the simulation time to zero. We performed two runs -- a control run, in which the cloud is simply left to its own devices, and a feedback run in which we placed a source of $10^{49}$ ionising photons s$^{-1}$ at the position $[-20, 0, 0]$ (measured in pc). We did not place a sink--particle at this location, since it would move under the influence of the cloud's gravity. Instead, ionising photons were simply taken to emerge spherically--symmetrically from this point throughout the simulation.\\
\indent We allowed the two clouds to evolve for a further $0.5$ freefall times ($2.4$ Myr). Our decision to terminate the runs at this point was motivated by two considerations. Firstly, the two simulations had clearly diverged sufficiently to allow an interesting comparison to be made. In addition, the ionising source we used represents a $\sim30$ M$_{\odot}$ star with a main--sequence lifetime of $\sim4\times10^{6}$ yr. The source could be expected to explode as a Type II supernova at the end of its main--sequence lifetime. To avoid this complication, we terminated our simulation well before the explosion is likely to occur.\\

\section{Results}

\subsection{Morphology of star formation}
\begin{figure*}
\includegraphics[width=\textwidth]{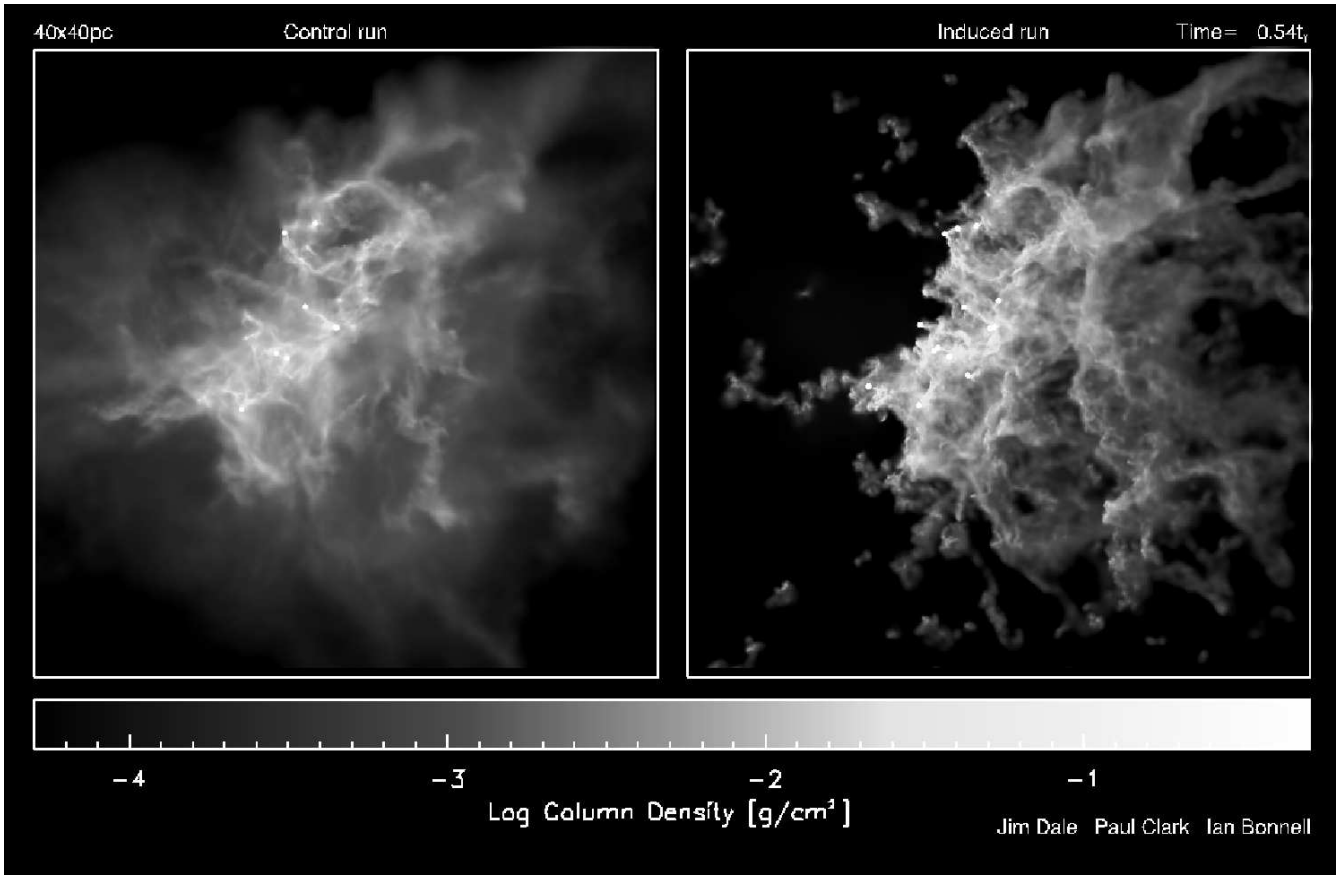}
\caption{Comparison of column--density maps (viewed down the z--axis) of the control run (left panel) and the feedback run (right panel) $0.5$ freefall times ($\sim2.4$ Myr) after ignition of the radiation source. The source is located halfway up the left--hand border of the right panel. White dots represent sink particles.}
\label{fig:turb_snap}
\end{figure*}

\begin{figure*}
\includegraphics[width=\textwidth]{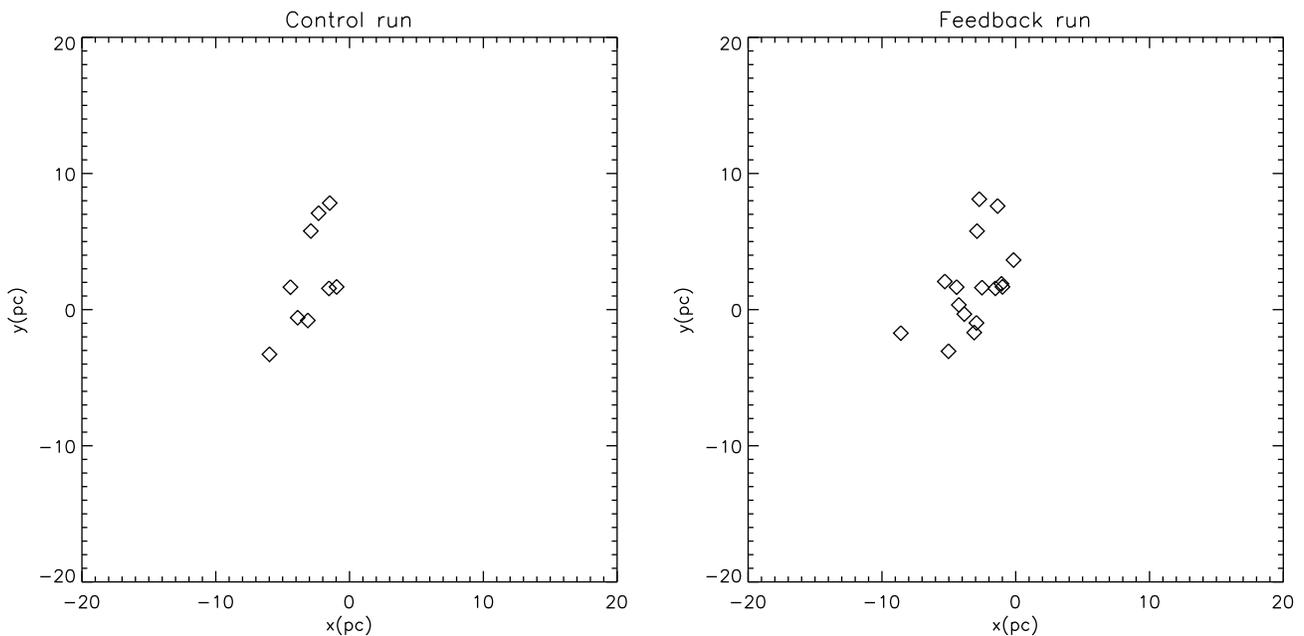}
\caption{Star formation sites in the control run (left panel) and the feedback run (right panel). The diamonds represent the position where each sink particle in the runs originally formed.}
\label{fig:sf_sites}
\end{figure*}

\begin{figure*}
\includegraphics[width=\textwidth]{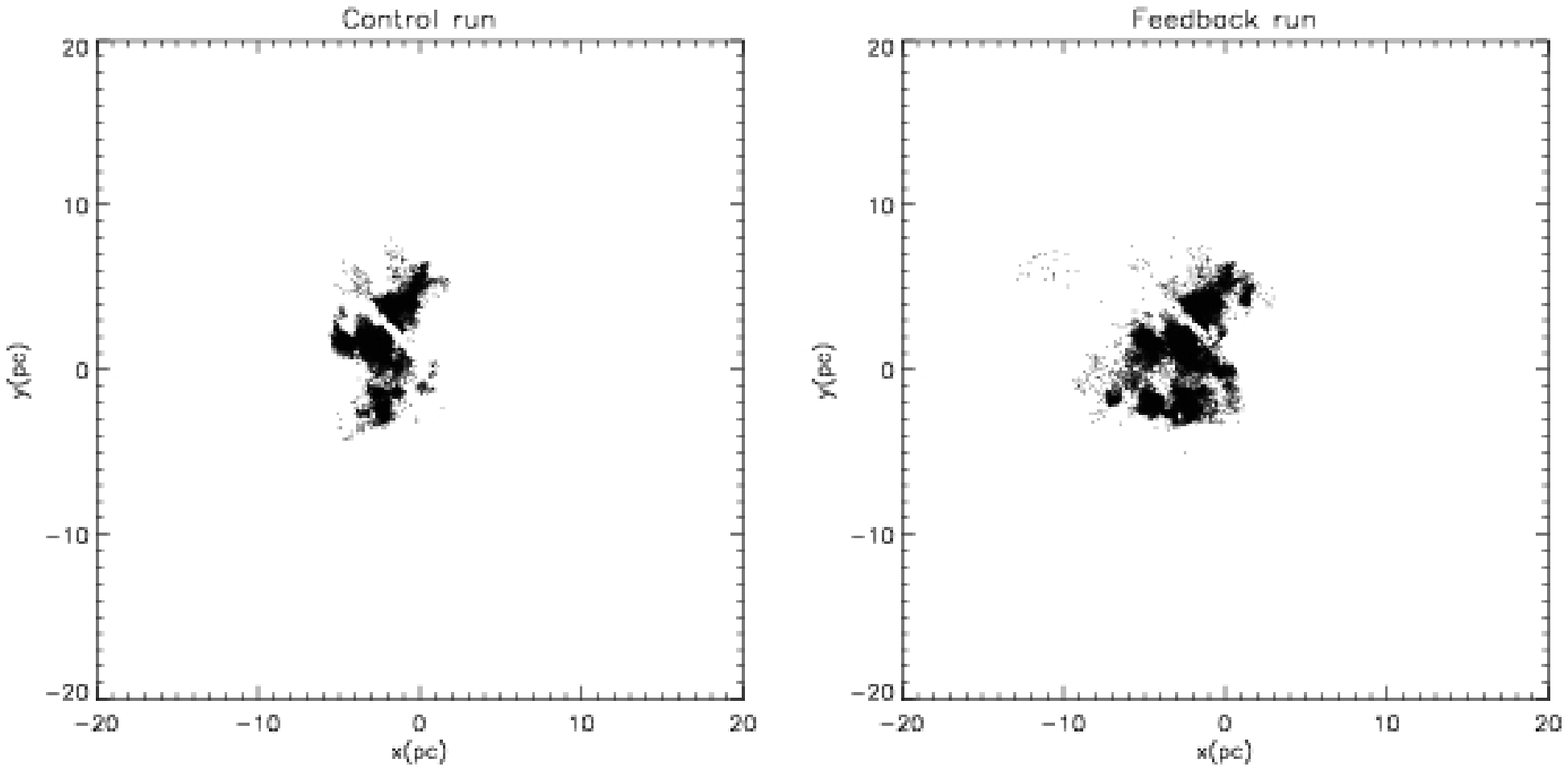}
\caption{Initial locations of all gas particles which formed part of a sink particle after $0.5$ freefall times ($\sim2.4$ Myr) in the control run (left panel) and in the feedback run (right panel).}
\label{fig:acc_sites}
\end{figure*}

In Figure \ref{fig:turb_snap} we compare column--density maps, as viewed along the z--axis of the control and feedback runs $0.5$ freefall times after ignition of the ionising source. Feedback has clearly had a dramatic effect on the morphology of the cloud. Most of the left--hand half of the cloud in the feedback run has been destroyed by the ionising radiation. It is also clear that more sink--particles have formed in the feedback run, and that they appear to be concentrated near the edge of the cloud exposed to the radiation, as might be expected if their formation has been triggered by feedback.\\
\indent We plot the sites of star formation in the two calculations in Figure \ref{fig:sf_sites}. The Figure shows that, although more star--forming cores form in the feedback simulation, star formation is confined to a similar volume in both calculations. However, Figure \ref{fig:acc_sites}, in which we plot the initial positions of all the gas particles that formed sink particles or were accreted by them, shows that the reservoir of gas from which the sink particles form in the feedback run is considerably more extended. This extension points approximately in the direction of the radiation source. The extra material has clearly been swept up by the expanding HII region and transported into the core of the cloud before either forging new star--forming cores or being accreted by pre--existing ones.\\
\subsection{Star formation efficiency and star--forming potential}
\begin{figure}
\includegraphics[width=0.5\textwidth]{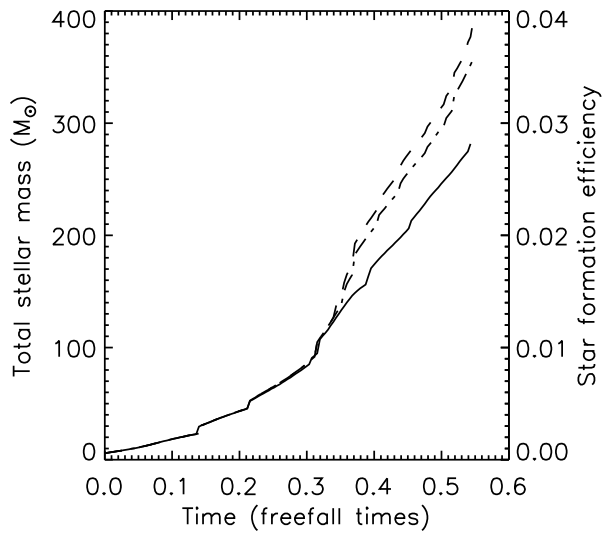}
\caption{Plot of total stellar mass and fractional star formation efficiency in the control run (solid line), feedback run (dashed line), and in the feedback run, excluding cores whose formation is triggered (dash--dot line).}
\label{fig:sfe}
\end{figure}

\begin{figure}
\includegraphics[width=0.5\textwidth]{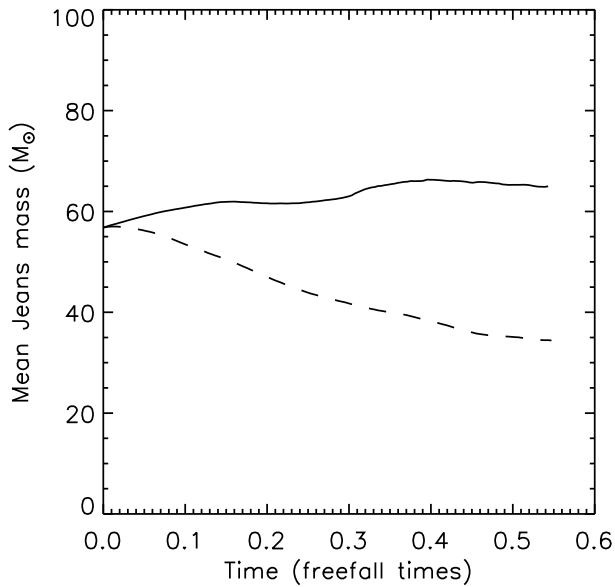}
\caption{Plot of the mean Jeans mass in the control run (solid line) and feedback run (dashed line). Only neutral gas is considered in the feedback calculation.}
\label{fig:jeans_mass}
\end{figure}

\begin{figure}
\includegraphics[width=0.5\textwidth]{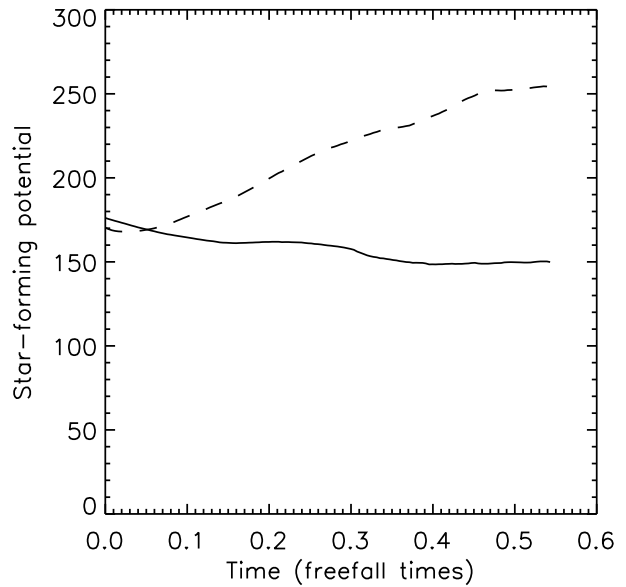}
\caption{Plot of the evolution of the Star Forming Potential in the control run (solid line) and in the feedback run (dashed line).}
\label{fig:sfp}
\end{figure}

\indent In Figure \ref{fig:sfe} we plot the evolution with time of the total stellar mass and fractional star--formation efficiency in the control and feedback runs. For the feedback run, we also include a plot of the stellar mass and star--formation efficiency excluding those cores that we later identify as having been triggered. We observe that the evolution of the two simulations in this sense is indistinguishable for $\sim0.3$ freefall times ($\sim1.4$ Myr) after ignition of the radiation source in the feedback run. This is a general feature of the star--formation process in these two calculations. Two conclusions can be drawn from this delay in divergence. Firstly, the material swept up by the HII region does not become involved in star--formation immediately. The swept--up material does not appear to become dense enough to collapse until it is driven into the denser undisturbed material towards the centre of the cloud, as seen by comparing Figures \ref{fig:sf_sites} and \ref{fig:acc_sites}. Secondly, and more obviously, the shock driven by the HII region cannot affect star formation \textit{which is already underway} immediately. In the control run, star formation takes place in the denser core of the cloud. In the feedback calculation, it takes considerable time for the shock driven by the ionisation front to penetrate this region.\\
\indent After $\sim0.3$ freefall times ($\sim1.4$ Myr), the rate at which the total stellar mass increases in the feedback calculation becomes larger than in the control run, so that after $0.5$ freefall times ($2.4$ Myr), star--formation in the feedback run has been a factor of about one third more efficient. As we show later, this is partly due to the fact that accretion rates onto the star--forming cores in the feedback run are generally higher, but also to the fact that more cores form in the feedback run -- after $0.5$ freefall times ($2.4$ Myr), the feedback run has formed $16$ cores, as opposed to $9$ in the control run. We also note that the triggered cores themselves make a relatively small contribution to the total stellar mass in the feedback run.\\
\indent Previous work by \cite{2005MNRAS.358..291D} showed that one effect of photoionising feedback on molecular clouds is to drive down the mean Jeans mass by compressing neutral gas in the clouds. In Figure \ref{fig:jeans_mass} we show that this effect occurs in this calculation too. We plot the evolution with time of the mean Jeans mass in the two calculations (considering only the neutral gas in the feedback run). We have used the mass--averaged density to calculate the Jeans mass, taking the mean density of all SPH excluding those whose densities exceed the threshold for sink--particle creation.\\
\indent The mean Jeans mass in the control run remains approximately constant. Although the cloud is unbound and therefore expanding, some of the inner regions are contracting (and engaged in star formation). These two effects roughly cancel each other. Since the calculation is isothermal, the mean Jeans mass is only dependent on the mean density and thus changes little over the course of the simulation. In the feedback run, by contrast, the mean Jeans mass falls almost linearly for the duration of the run. This is due to low--density neutral gas being swept up by the expanding HII region, as also seen in the calculations of \cite{2005MNRAS.358..291D}.\\
\indent A drop in the mean Jeans mass should encourage fragmentation of neutral gas which one might naively think would accelerate the star formation process. To quantify this idea, we define the `\textit{star--forming potential}' (SFP) of a molecular cloud as the number of stars a cloud is likely to form in the future. We estimate the SFP by
\begin{eqnarray}
SFP(t)=\frac{M_{ntrl}(t)}{\left<M_{J}(t)\right>},
\end{eqnarray}
where $M_{ntrl}$ is the total quantity of neutral gas in the cloud and $\left<M_{J}(t)\right>$ is the mean Jeans mass, and we have explicitly emphasised the time--dependence of each quantity. This estimate is clearly crude - not all the neutral gas in any given cloud is likely to form stars, and denser material is likely to fragment at a mass less than the mean Jeans mass. However, the SFP still gives a rough idea of how many stars a molecular cloud might be expected to form. Strictly, one would have to assume that the neutral mass and mean Jeans mass are changing on a timescale longer than the timescale on which star formation is proceeding, but the SFP is not intended to be a rigorous measure, so we neglect this complication.\\
\indent In Figure \ref{fig:sfp} we plot the evolution of the SFP as defined above in the control and feedback runs. In the control run, the quantity of neutral gas available (in principle) for forming stars is constant, so the changes in the SFP simply mirror the changes in the mean Jeans mass shown in Figure \ref{fig:jeans_mass}. In the feedback run, neutral gas is continuously being ionised and blown away from the cloud, so the behaviour of the SFP in this calculation is more complex. The SFP initially falls below that in the control run, as a large quantity of neutral gas is rapidly ionised and blown off the surface of the cloud by the sudden ignition of the ionising source. As the shock driven by the HII region sweeps up and compresses the neutral gas in the feedback run, decreasing the mean Jeans mass, the SFP quickly climbs above that in the control run. It appears to stabilise towards the end of the calculation at a value $\sim60\%$ higher than in the control run. We note, however, that neither simulation comes close to achieving its star--forming potential during our simulations (which have a duration of $\sim0.5$ freefall times.\\
\indent Given that the SFP in the feedback run is greater for almost the whole duration of the simulation, it is perhaps not surprising that this calculation produces more star--forming cores. However, we have not definitively answered the question of whether feedback is inducing star formation in the strong sense outlined in the Introduction, or merely accelerating the formation of unstable cores that would have eventually formed anyway. To do this, we must compare the histories of the sink particles and the gas from which they form in the two calculations. Fortunately, the Lagrangian nature of SPH makes it ideal for this task.\\
\subsection{Histories of individual cores and of star--forming gas}
Watching animations of the two simulations side--by--side gives the strong impression that some sink particles are `the same' in both calculations -- they form at approximately the same times and places. As the calculations progress and diverge from each other, such identifications become much more difficult. To analyse this possibility objectively, we traced the history of every particle in each simulation which was involved in the \textit{formation} of a sink particle (as opposed to being accreted by a pre--existing one). Since the initial conditions of the calculations are identical, we can trace the history of these particles in both calculations and determine which sinks form in both calculations and which (if any) sinks are induced to form, or prevented from forming, in the feedback calculation. If most or all (we used a criterion of $>75\%$) of a given group of particles which forms a sink particle in one calculation also forms a sink in the other, we consider those sink particles to be the same object. If a group of particles becomes a sink particle in one calculation, but not the other, we examine the evolution with the time of the group's thermal, kinetic (in the group's centre--of--mass frame) and gravitational energy in an attempt to infer the fate of the group. There are four possibilities:\\
(1) gravitational energy becoming more negative and kinetic energy decreasing -- group contracting and will probably form a sink eventually\\
(2) gravitational energy become more negative and kinetic energy increasing -- group collapsing and will probably form a sink eventually\\
(3) gravitational energy becoming less negative and kinetic energy decreasing -- group expanding but expansion slowing, so group may form a sink later\\
(4) gravitational energy becoming less negative and kinetic energy increasing -- group expanding and expansion accelerating, so group is unlikely to form a sink\\
Using this technique, we performed a census of all the star--forming cores formed in each simulation and cross--correlated them. We present the results in Tables \ref{tab:control} and \ref{tab:induced}. For convenience, we have grouped the cores formed in each run according to what happened to those same cores in the counterpart run, i.e. whether they formed earlier, later or not at all. We give each group of cores a `population identifier' -- populations from the control run are prefixed with a `C' and those from the feedback run with an `F'. For the material forming a given core in a given run, there are five possible outcomes for that same material in the counterpart run. If it forms the same object at the same time, we identify the core as a member of population C1 in the control run and of population F1 in the feedback run. Feedback had no effect on the formation times of these objects. Cores in population C2 in the control run are objects that form earlier in the feedback run as population F2. Conversely, the cores in population C3 in the control run are objects whose counterparts in the feedback run form later, and are labelled population F3. Here, earlier or later means a difference in formation time of at least $1.5\times10^{4}$ yr. Population C4 consists of star--forming cores which form in the control run and not in feedback run, but whose formation in the feedback run in the future appears likely. Similarly, objects in population F4 have formed in the feedback run and will probably form later in the control run. Of most interest are the final two populations. Population C5 are cores which are highly unlikely to form in the feedback run -- their formation has been aborted or prevented, so they are examples of negative feedback. In contrast, population F5 are highly unlikely to form in the control run. These cores have been triggered in the strong sense. We now examine these populations in more detail.\\

\begin{table*}
\begin{tabular}{|l|l|l|l|}
\hline
Population identifier & Fate of cores in feedback run & Number of cores & Fraction of total (9)\\
\hline
C1 (=F1) & Form at same time & 2 & 0.22 \\
C2 (=F2) & Form earlier & 4 & 0.44\\
C3 (=F3) & Form later & 2 & 0.22\\
C4 & May form later & 1 & 0.11\\
C5 & Do not form & 0 & 0.00\\
\hline
\end{tabular}
\caption{Fate of cores \textit{formed in the control run} in the feedback run.}
\label{tab:control}
\end{table*}

\begin{table*}
\begin{center}
\begin{tabular}{|l|l|l|l|}
\hline
Population identifier & Fate of cores in control run & Number of cores & Fraction of total (16)\\
\hline
F1 (=C1) & Form at same time & 2 & 0.13 \\
F2 (=C2) & Form later & 4 & 0.25\\
F3 (=C3) & Form earlier & 2 & 0.13\\
F4 & May form later & 5 & 0.31\\
F5 & Do not form & 3 & 0.19\\
\hline
\end{tabular}
\caption{Fate of cores \textit{formed in the feedback run} in the control run.}
\label{tab:induced}
\end{center}
\end{table*}

\begin{figure*}
\includegraphics[width=\textwidth]{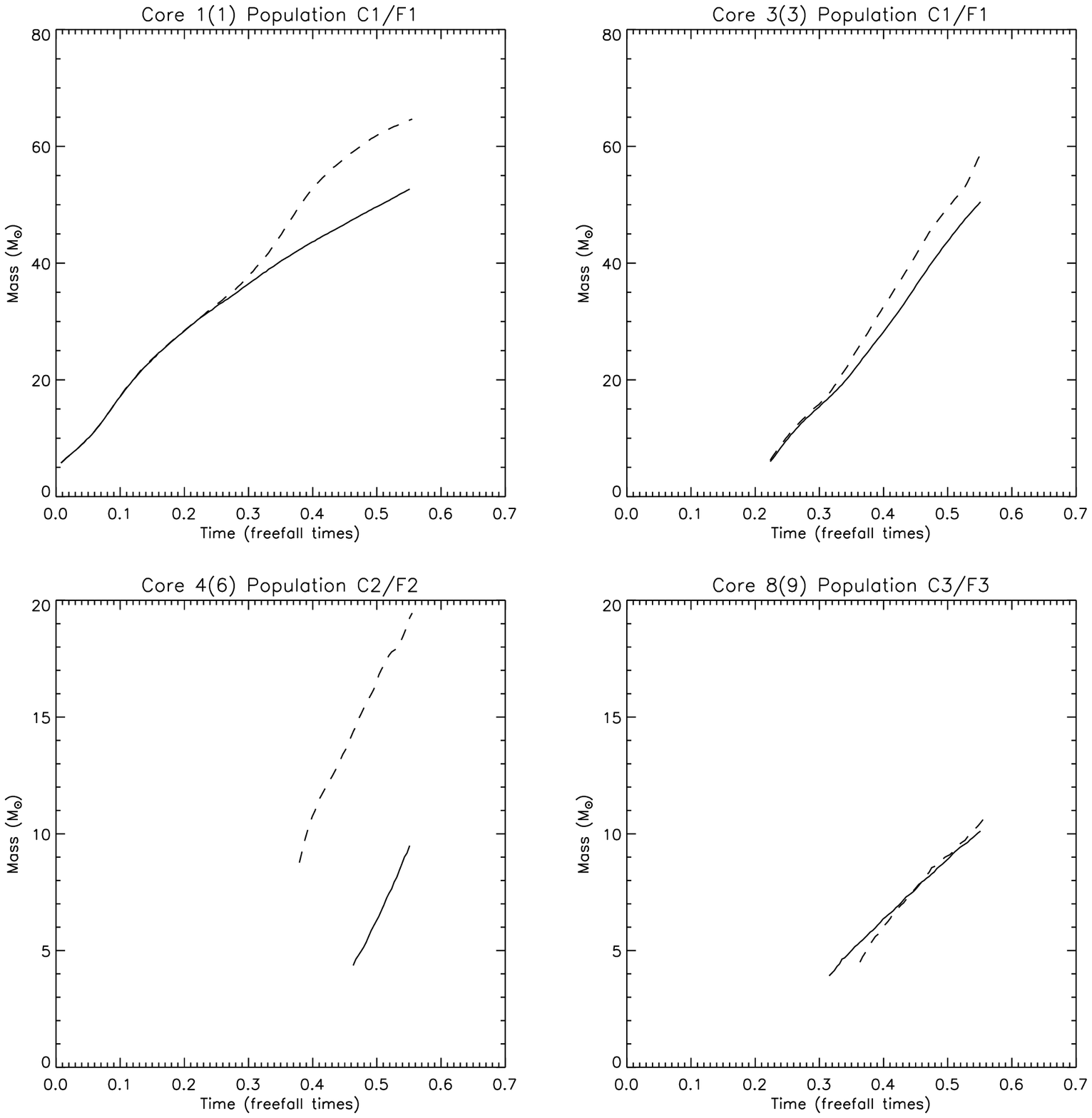}
\caption{Accretion histories of selected objects from populations C1, C2 and C3 (solid lines), and F1, F2 and F3 (dashed lines). Numbers assigned to cores are the order in which they formed (order in the feedback run in parentheses), and the populations to which they belong are indicated.}
\label{fig:acc_c1f1}
\end{figure*}

\indent \textbf{Cores which form in both calculations (populations C1=F1, C2=F2 and C3=F3)}\\
In Figures \ref{fig:acc_c1f1}, we plot the increase in mass with time for two cores in the C1/F1 populations, one core from the C2/F2 population and one from the C3/F3 population. We interpret population C1/F1 as cores whose formation times are unaffected by feedback. However, even though these cores form at the same time in the two simulations, Figure \ref{fig:acc_c1f1} shows that the two accretion histories of each core are different. The accretion histories of the C1/F1 populations are the same until $\sim0.3$ freefall times ($1.4$ Myr) after ignition of the ionising source, after which the cores in the F1 population generally accrete mass more quickly than their counterparts in the control run. Cores in populations C2/F2 and C3/F3 also experience accretion rates in the feedback run greeater than or equal their accretion rates in the control run. This is due to the increased gas density in the cloud core resulting from compression caused by the expanding HII region.\\
\indent The cores belonging to population C2 are induced in the weak sense outlined in the introduction -- they would have formed anyway, but they have been induced to form earlier by the action of feedback. Those in population C3 hint that negative feedback is also occurring, since their formation has been delayed. The three populations C1/F1, C2/F2 and C3/F3 are examples of revealed star formation. Eventually, the gas around these objects will be washed away by the ionisation front, leaving them isolated. Their proximity to the ionisation front and the absence of any information on their formation history might lead one to conclude that their formation had been triggered. It is only by comparison between our two calculations that we can say with certainty that they would in fact have formed spontaneously in the absence of feedback.\\
\indent \textbf{Cores which form in the control run but not in the feedback run (population C5)}\\
These are cores whose formation is \textit{aborted} by feedback because the group of particles from which they formed in control run are disrupted. We do not observe any objects of this nature in this calculation.\\

\indent \textbf{Cores which form in the feedback run but not in the control run (population F5)}\\
\begin{figure*}
\includegraphics[width=\textwidth]{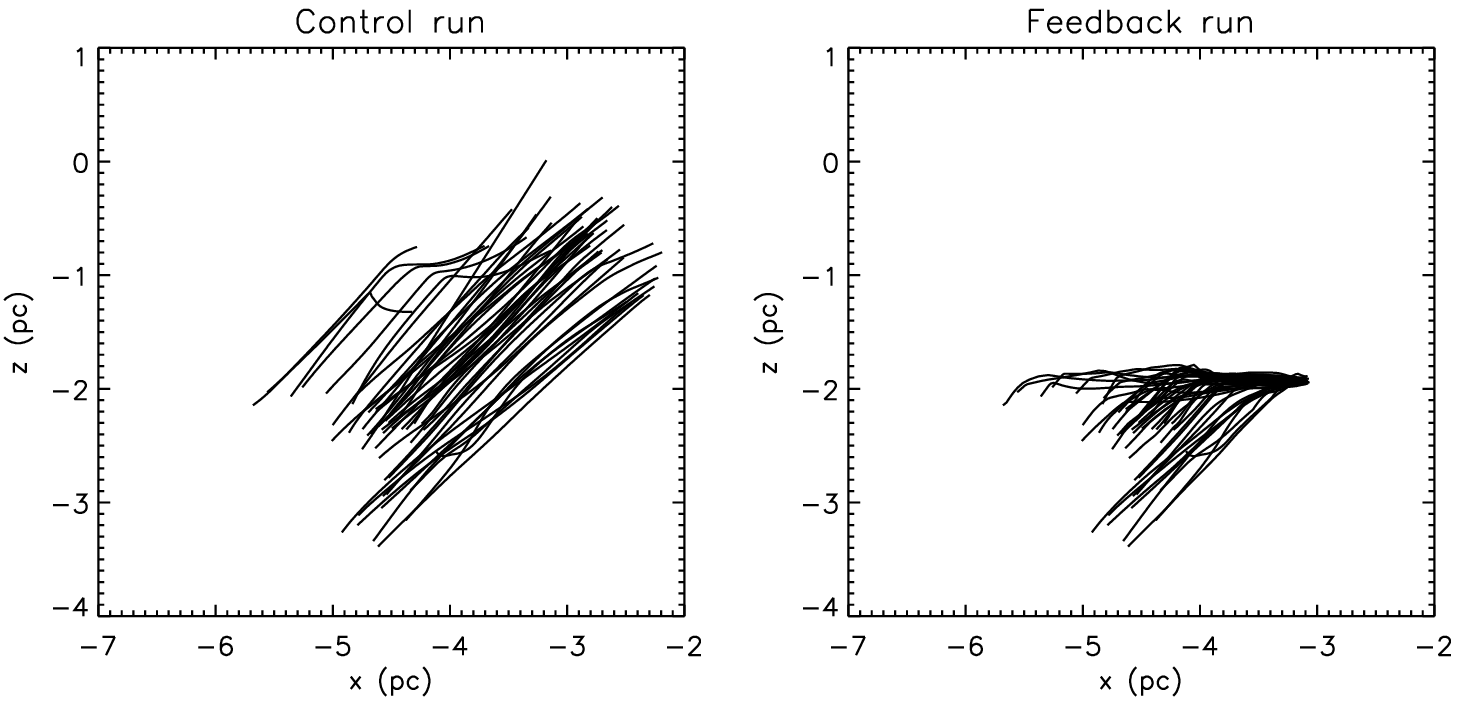}
\caption{Tracks of the SPH gas particles forming one of the triggered cores, shown in the control run (left panel) and feedback run (right panel). Particle motion is initially approximately from bottom left to top right. In the control run, this group of particles is dispersed by fluid flows. In the feedback run, the group encounters a shock by which it is compressed, resulting in the formation of a star--forming core.}
\label{fig:triggered_core}
\end{figure*}
These are cores whose formation is induced in the strong sense -- we can say with a high degree of confidence that they would not have formed in the absence of feedback. An example is shown in Figure \ref{fig:triggered_core}. The existence of these objects demonstrates that feedback can potentially increase the final star--formation efficiency of a molecular cloud. To quantify the increase, the two clouds would have to be evolved until they both cease forming stars, perhaps after internal supernova explosions expel their remaining gas. However, we do not attempt this here.\\
\indent We do not discuss groups C4 or F4, the cores which form in one run and may still form in the other. While it is true that evolving the two simulations further would reveal the fate of some of these objects, there will always be some star--forming cores for which the fate of their counterparts in the companion simulation is uncertain. Since the simulations have run far enough to provide us with examples of revealed, delayed and triggered star formation (in both strong and weak senses), we do not attempt to follow the evolution of these objects any further.\\

\section{Observable outcomes of triggered star formation}
In this work, we have performed a numerical experiment to determine whether external irradiation of a turbulent molecular cloud can trigger star formation within the cloud. We have the benefit not only of being able to examine and replay the history of our model feedback--influenced molecular cloud as many times as we please, but also of being able to compare it with a control simulation in which feedback is absent. Observers, unfortunately, have neither of these advantages, since they only see a single snapshot of any given system. In addition, any real observation is contaminated with foreground and background sources and extincted by intervening dust, and does not come with full spatial and kinematic information attached.\\
\indent We have identified star--forming cores in our feedback simulation which we can say with a high degree of confidence would not have formed in the absence of feedback. We were able to do this by comparing the histories of the gas particles from which these populations formed with their counterparts in the control run. Of particular interest is whether there is any potentially--observable characteristic that differentiates the triggered F5 stellar population from the other populations in the feedback run. If such a characteristic exists which allows the induced stars to be picked out from a snapshot of the simulation, this would be of enormous help to observers studying a similar system. We therefore compared the properties of the induced cores with those of their colleagues in the feedback run to see if there was anything to make them stand out. Obviously, conclusions in this section should be treated with caution as they rely on very small numbers of objects.\\
\indent Several other studies of induced star formation (e.g. \cite{1994A&A...290..421W}) have suggested that stars whose formation has been triggered should be of higher mass than those which form unassisted. This is not the case in this simulation. The rate at which a given star or star--forming core accretes mass and the final mass which it achieves depends on its time--integrated local gas density. Accretion rates in the feedback run are higher than in the control run, but this applies to \textit{all} objects. There is therefore nothing about the final masses of the cores which distinguishes the induced ones from their colleagues. In fact, the mean core mass after $0.5$ freefall times ($\sim2.4$ Myr) is $\sim31$ M$_{\odot}$ in the control run and $\sim24$ M$_{\odot}$ in the feedback run, so feedback has \textit{lowered} the mean core mass. The reason for this is that, although both simulations have a few massive cores, those that form early on and have accreted large quantities of mass, the feedback run has a larger population of recently--formed and therefore lower mass objects.\\
\indent Stellar ages are also often used to infer whether a population of stars has been triggered. The presence of very young stars near a massive star is certainly suggestive. We examined the feedback run to see if there existed a gradient in core age (and, by assumption, stellar age) with distance from the ionising source but did not find this. Because of the complex morphology of the molecular cloud, even before the ignition of the O--star, there is dense and potentially star--forming material located at a variety of distances from the ionising source which can be influenced by the ionisation front at roughly the same time. One might actually expect there to be a weak negative correlation of stellar age with distance from the source, since the ionisation and shock fronts take time to penetrate the denser regions of the cloud, but we do not observe this either.\\
\indent Since the formation of the induced cores is triggered by the ionisation front, it might be thought that they should move along with the front, or at least to have velocity vectors of relatively large magnitude directed away from the ionising source. This is not true of the induced cores in this simulation. The velocity field of the star--forming cores in both the feedback and control runs appear to be largely a consequence of the initial turbulent velocity field and of the general expansion of the cloud. Neither the velocities, momenta or kinetic energies of the induced cores serve to pick them out. We also checked to see whether the induced cores could be identified by their proximity to the ionisation front, but they cannot. In the feedback calculation, as can be guessed from the right panel of Figure \ref{fig:turb_snap}, most of the star--forming cores, regardless of whether they were triggered or not, are located near the ionisation front, since the front sweeps up denser clumps of gas which are already in the process of forming cores, along with gas which would not otherwise by star--forming.\\
\indent The SPH code used in these calculations records one further property of sink particles, namely their spin angular momentum (derived from the relative angular momentum of the gas particles from which the sink originally formed and from the angular momentum of gas particles it accreted later). There is no obvious reason why the spins of the induced cores should be unusual, but we examined them nonetheless. Again, there is nothing to make the induced cores stand out.\\
\indent We therefore exhausted the physical properties that could potentially be used to flag the induced cores. Of course, the statistics on which these conclusions are based are very poor, relying on nine sink particles in the control run and sixteen in the feedback run. It is possible that larger populations of star--forming cores would reveal a statistical correlation between some physical property and whether or not a given core was induced to form. One might expect, for example, that the core mass functions of the two runs would be different, but we have not attempted to construct mass functions for our very small populations. Further numerical work at higher resolution and utilising more powerful computers will be required to study this possibility in detail.

\section{Conclusions}
We have demonstrated that external irradiation of a GMC by an O--type star is able to increase the star--formation efficiency of molecular clouds. The answers to all the questions posed in the Introduction are affirmative:\\
(i) ionising feedback can accelerate the formation of star--forming cores that would have formed anyway.\\
(ii) feedback can delay or the formation of cores, although we did not observe the disruption of  the any cores in our feedback simulation.\\
(iii) feedback can cause the birth of stars that would not otherwise form.\\
\indent Feedback can thus trigger star formation in both the strong sense of forming extra stars and in the weak sense of merely accelerating the formation of stars that were forming anyway. In the simulations presented here, these effects dominate over the destructive effects of feedback. We have therefore shown that feedback \textit{can} increase the star formation efficiency of a molecular cloud in the sense of causing it to form more stars overall than it would otherwise. In these simulations, the overall effect is rather small, increasing the star--formation efficiency by only $\sim30\%$ over the course of $\sim0.5$ freefall times. However, we have not shown that feedback always has this result. The effect of ionising radiation on molecular clouds is a complex interplay between the competing effects of destructive ionisation and ejection of neutral material, and the compression of molecular gas and consequent encouragement of fragmentation and star--formation. Further simulations are required to determine what properties of molecular clouds determine which of these effects will win. We have only examined a single point in parameter space and it is likely that some molecular clouds with parameters different from ours would experience strong negative feedback.\\
\indent The observational implications of this work are less clear. All the analysis we have performed is comparative -- we have compared the results of our feedback run with those of our control run. We examined the masses, velocity components, total velocity, rotation components, total spin, momentum components, total momentum and kinetic energy of our star--forming cores to see if there were any relations that clearly picked out the induced cores in the feedback run. Unfortunately, we found nothing that made the induced cores stand out. They appear to be indistinguishable from their untriggered colleagues. This finding may well be due to the extremely poor statistics of our calculations. Simulations of the kind presented here but which form a much larger number of objects may reveal two (or more) observationally distinguishable populations that can be labelled triggered and untriggered without reference to a control run.\\

\section{Acknowledgments}
We thank an anonymous referee whose comments and suggestions improved the paper considerably. JED acknowledges support from the University of Leicester's PPARC rolling grant. 

\bibliography{myrefs}

\label{lastpage}

\end{document}